\documentclass{pos}

\title{Contributions to the QCD Pressure Beyond Perturbation Theory}

\ShortTitle{QCD Pressure Beyond Perturbation Theory}

\author{\speaker{Klaus Lichtenegger}\\
        Institut f\"ur Physik, Graz University\\
        E-mail: \email{klaus.lichtenegger@uni-graz.at}}

\author{Daniel Zwanziger\\
        Department of Physics, New York University\\
        E-mail: \email{daniel.zwanziger@nyu.edu}}

\abstract{In this article we report on a new proposal to treat
the infrared problems of thermal QCD by taking into account
explicitly the confining influence of the Gribov horizon.
In order to make clear the possible value of such an approach,
we briefly review the most important arguments why a
straightforward perturbative description of finite-temperature
QCD is unlikely to be successful. From the infrared problems
of thermal perturbation theory one can conclude that confinement
effects and bound states probably play an important role also
in the high-temperature phase.

To set the stage we recount the supposed role of the
Gribov horizon for confinement, before we turn to the application
to finite-temperature theory. In the current approach it has been
found that the contributions to the free energy from the explicit
inclusion of the horizon begin to set in at order $g^6$ -- precisely
where the infrared problems of thermal QCD lead to a
breakdown of ordinary perturbation theory.

From the study of observables (free energy, anomaly, bulk viscosity)
we also note that for thermodynamic observables the leading order
term obtained by such an expansion in the coupling strongly deviates
from the more complete numerical solution. This can be regarded as
yet another sign for general problems of series expansions in
thermal QCD.

}

\FullConference{8th Conference Quark Confinement and the Hadron Spectrum\\
                 September 1-6, 2008\\
                 Mainz. Germany}

\begin{document}

\section{Infrared Problems in Thermal QCD}

The properties of QCD at high temperatures ($\mathcal{O}(100\,\mathrm{MeV})$
and higher) are obviously  important for an understanding of the early universe.
While this domain is to some extent accessible experimentally (at RHIC and
LHC), it remains a challenge for theory.

Initially one had expected to find a phase transition where at some temperature
hadrons melt into a ``quark-gluon plasma''~\cite{Shuryak:1978ij}. In such
a plasma, quarks and gluons were supposed to be almost free particles
which could be reliably described by perturbation theory. Such a purely
perturbative description was motivated by the fact that high temperatures
imply high average momentum transfer $\mu$ and thus -- due to asymptotic
freedom -- a small coupling $g(\mu)$.

However, this picture turned out to be too naive. In principle, this should have
been clear at least since 1980, when it was shown~\cite{Linde:1980ts, Gross:1980br}
that at order $g^6$ a natural barrier arises for any perturbative description (the
\emph{Linde problem}). Even earlier than that, the simple fact that the
infinite-temperature limit of four-dimensional Yang-Mills theory is a
three-dimensional \emph{confining} Yang-Mills theory could and should
have been regarded as a sign that any straightforward perturbative approach
to high-temperature QCD was problematic.

Still it took more than 20 years until it began to be accepted
that the high-temperature phase of QCD has little to do with a
conventional plasma. The results of the RHIC experiments~\cite{Shuryak:2004cy}
clearly showed that also above the transition, bound state phenomena
can not be neglected.

\smallskip

Since perturbation theory is limited to a finite order (and, in addition,
for experimentally accessible temperatures the apparant convergence
of the perturbation series is bad~\cite{Kajantie:2002wa, Blaizot:2003iq}),
other methods have been invoked to treat the properties of finite
temperature QCD:

One popular proposal~\cite{Braaten:1995cm} is to explicitly separate the
three relevant scales (hard -- $2\pi T$, chromoelectric -- $gT$ and
chromomagnetic -- $g^2T$) by introducing effective theories at each scale.
Two of these theories can be treated perturbatively, while one (which governs
the magnetostatic sector) is genuinely nonperturbative.

Therefore nonperturbative methods are -- either in the direct or in the effective theory
approach -- urgently needed. One possibility is given by functional methods,
based for example on Dyson-Schwinger equations; for an application to finite
temperature see e.g.~\cite{Maas:2004se, Maas:2005hs}.
These methods provide valuable insights, but unfortunately up to now
the pressure (and quantities derived from it) are difficult to access. Still
functional methods provide further evidence for the picture of bound states
playing an important role even at very high temperatures and parts of the
gluon spectrum being confined at any temperature.

The nonperturbative methods most actively pursued for gauge theories are
lattice-based, and indeed lattice data is available for finite temperatures up to
$T\approx 5 T_C$. In this region for pure SU(3) gauge theory the problem of
determining the equation of state is regarded as solved~\cite{Boyd:1996bx}.
Also the AdS/CFT duality~\cite{Maldacena:1997re} and the AdS/QCD
duality~\cite{Aharony:1999ti} are employed to study
finite temperature gauge theories.

In this article we report on a project which follows an alternative path:
Since the footprints of confinement are visible also in the ``deconfined'' phase,
the confining influence of the Gribov horizon is explicitly taken into account.
The method  has been proposed in~\cite{Zwanziger:2006sc, Zwanziger:2007zz}
and further developed in~\cite{Zwanziger:2006wz, Lichtenegger:2008mh}.
To explain it, we will briefly summarize the importance of the Gribov horizon
in section~\ref{sec:kldz_gribovhorizon}, while in section~\ref{sec:kldz_appfintemp}
we will discuss the application to finite-temperature theory.

\newpage

%%%%%%%%%%%%%%%%%%%%%%%%%%%%%%%%%%%%%
\section{The Gribov Horizon and Confinement}
\label{sec:kldz_gribovhorizon}

\begin{figure}
  \includegraphics[width=7cm]{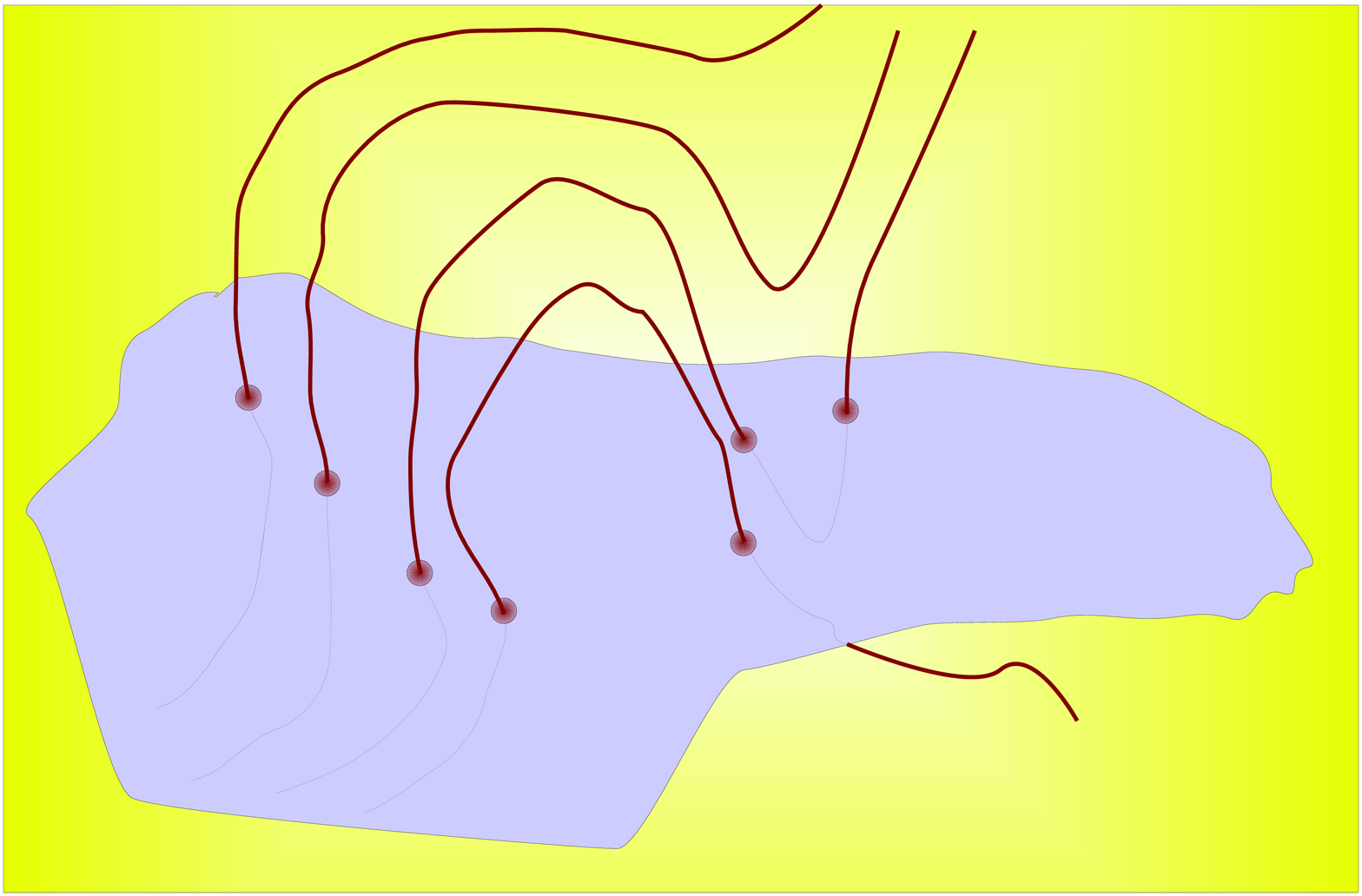}
  \put(-196,120){(a)}\put(-48,82){$F[A]=0$}
  \put(-173,114){\small $O_1$} \put(-148,100){\small $O_2$}
  \put(-65,96){\small $O_3$} \put(-118,92){\small $O_4$}
  \put(-176,74){\scriptsize $A_1^{(0)}$} \put(-164,62){\scriptsize $A_2^{(0)}$}
  \put(-150,50){\scriptsize $A_3^{(0)}$} \put(-132,48){\scriptsize $A_4^{(0)}$}
  \put(-72,70){\scriptsize $A_3^{(2)}$}
  \put(-89,60){\scriptsize $A_3^{(1)}$} \put(-104,52){\scriptsize $A_4^{(1)}$}
  \hspace{.8cm}
  \raisebox{1cm}{\includegraphics[width=7cm]{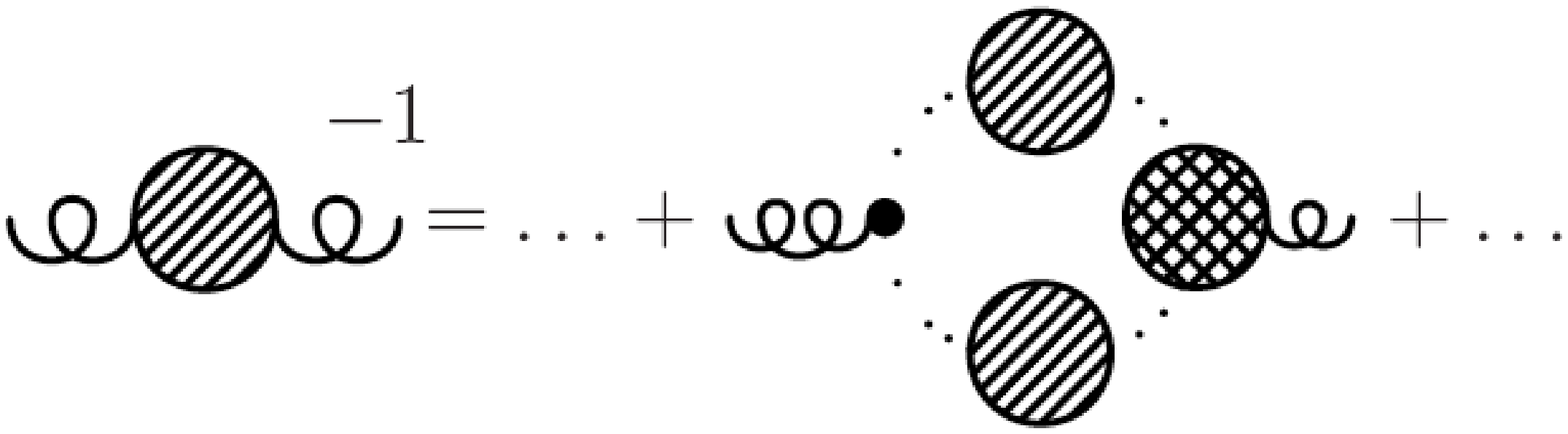}
  \put(-200,92){(b)}}
  \caption{(a) In non-Abelian gauge theories there are typically
  configurations $A_i^{(k)}$, $k=0,1,2,\dots$ on the same gauge orbit
  $O_i$ which fulfill the same gauge condition $F[A]=0$. (b) The
  gluon Dyson-Schwinger equation (only the infrared-dominant part
  is displayed on the rhs).}
  \label{fig:kldz_gribovcopies}
\end{figure}

The initial goal of gauge-fixing was to single out precisely one configuration
on each gauge orbit in order to embed the physical configuration space
into the space of all field configurations.
As it has been shown by Gribov~\cite{Gribov:1977wm} and further discussed
by Singer~\cite{Singer:1978dk}, in non-Abelian gauge theories the usual
gauge-fixing procedure does not yield uniqueness. So for any local gauge-fixing
condition $F[A]=0$ one typically finds, as sketched in
figure~\ref{fig:kldz_gribovcopies}a, several gauge-equivalent
configurations $A^{(0)}$, $A^{(1)}={}^{g_1}A^{(0)}:=g_1^{-1}A^{(0)}g_1 + g_1^{-1}\partial g_1$,
$A^{(2)}={}^{g_2}A^{(0)}$, \dots which all fulfill this condition.

This does not affect perturbation theory (since only the vicinity of $A=0$ plays
a role there), but in order to account for nonperturbative effects, one has to consider
the effect of such \emph{Gribov copies}. In particular the infrared properties
are dramatically changed.

In order to eliminate Gribov copies one typically restricts the domain
of integration to the \emph{Gribov region}, i.e. the region where the Faddeev-Popov
operator $\mathcal{M}(A)$ is positive semidefinite.\footnote{However, as
discussed in~\cite{vanBaal:1991zw}, in principle one has to restrict the
integration even further -- to the Fundamental Modular Region (FMR), which
is unfortunately characterized by a global condition and thus difficult to handle.}
The boundary of the Gribov region (where $\mathcal{M}(A)$ has at least one
zero eigenvalue) is called the \emph{Gribov horizon}.
The presence of the Gribov horizon (and the restriction of the functional
integral to the Gribov region) leads to modifications of the Faddeev-Popov
procedure~\cite{Zwanziger:1982na}. The influence of the Gribov horizon
can be incorporated in a local action~\cite{Zwanziger:1989mf, Zwanziger:1991gz}.

\smallskip

While the Gribov problem seems to be a nuisance at first glance, in
fact a whole confinement scenario is based on it -- proposed in~\cite{Gribov:1977wm}
and further elaborated by one of the authors~\cite{Zwanziger:1989mf, Zwanziger:1991gz}:
Since the Gribov region is (on the lattice) a high- or (in the continuum) an infinite-dimensional
region, due to geometrical reasons most of the volume is concentrated close to the 
boundary (``the entropy argument''). So most configurations which contribute to the path
integral have an almost vanishing Faddeev-Popov operator and correspondingly
an enhanced ghost propagator.

Such an enhanced ghost can be made responsible for the suppression of the
gluon propagator in the infrared. This is most easily understood by analyzing
the gluon Dyson-Schwinger equation
(see figure~\ref{fig:kldz_gribovcopies}b)~\cite{von Smekal:1997is,
von Smekal:1997vx, Alkofer:2000wg}. The ghost loop on the right-hand side is infrared
enhanced and thus the gluon propagator is accordingly suppressed -- the gluon cannot
propagate over long distances.
    
\newpage

%%%%%%%%%%%%%%%%%%%%%%%%%%%%%%%%%%%%%%%%%%%%%%
\section{Application to Finite Temperature Theory}
\label{sec:kldz_appfintemp}

We now turn to the ``marriage'' of finite temperature theory and Gribov's confinement
scenario. The basic physical idea is that the infrared divergences of finite-temperature
perturbation theory (which are responsible for the Linde problem) do not arise
when the domain of functional integration is cut off at the Gribov horizon.
The cut-off is done in Coulomb gauge which is well adapted to finite-temperature
calculations.

Technically the cut-off at the Gribov horizon is implemented by adding a
``horizon function'' to the action \cite{Zwanziger:1989mf, Zwanziger:1993}.
The initially non-local term then gets replaced by a local, renormalizable term
in the action by means of an integration over a multiplet of auxiliary ghost fields.
The new term in the action depends on a mass parameter $m$; the functional
cut-off at the Gribov horizon imposes the condition that the free energy $W$
or quantum effective action $\Gamma$ be stationary with respect to that mass,
$\frac{ \partial W }{ \partial m } = - \frac{ \partial \Gamma }{ \partial m } = 0$.

\begin{figure}
\begin{center}
  \hspace{.4cm}
  \includegraphics[width=7cm]{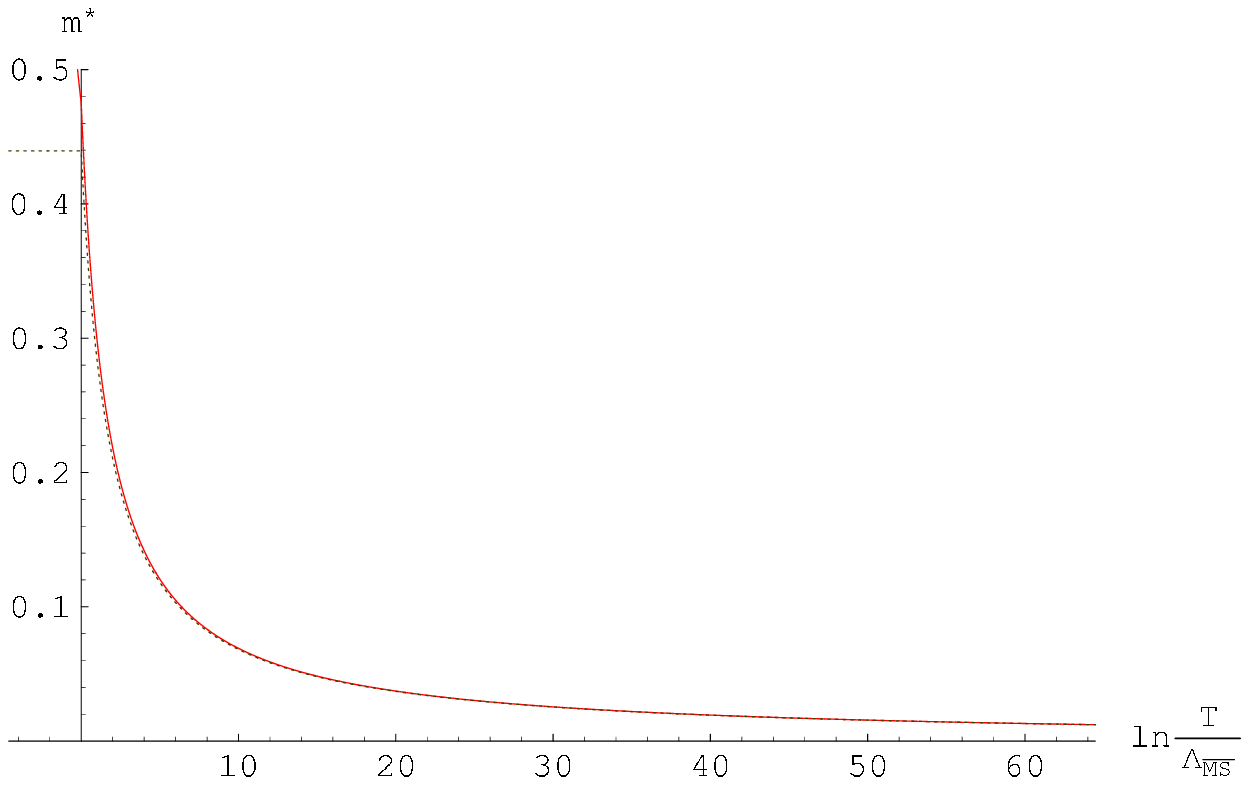} \put(-222,111){(a)}
  \hspace{.4cm}
  \includegraphics[width=7cm]{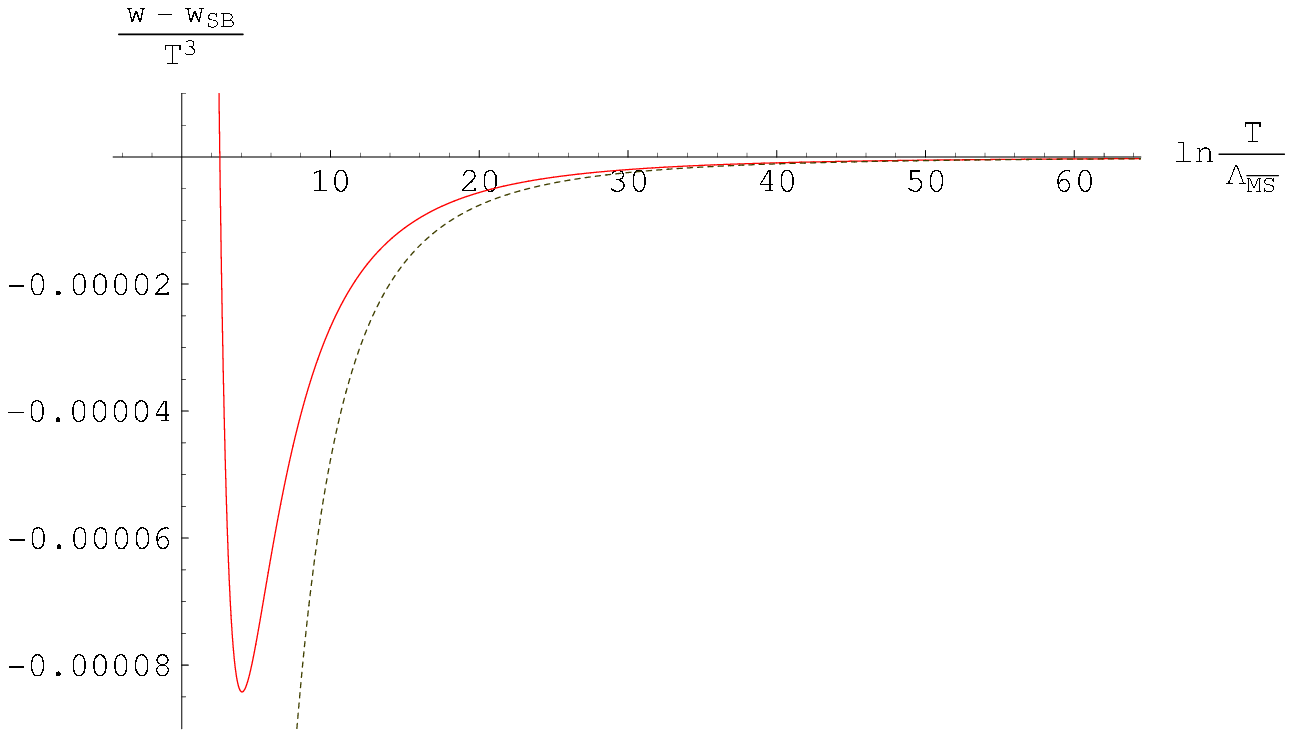} \put(-212,111){(b)}
\end{center}
\caption{(a) the rescaled Gribov mass $m^*=\frac{m}T$, (b) the
 rescaled free energy $\frac{w-w_{\mathrm{SB}}}{T^3}$, where
 the Stefan-Boltzmann term $w_{\mathrm{SB}}$ has been subtracted. [solid --
 numerical solution, dotted -- leading order expansion]}
\label{fig:kldz_massfreeenergy}
\end{figure}

This ``horizon condition" has the form of a non-perturbative gap equation
that determines the Gribov mass~$m = m(T, \Lambda_{QCD})$ and thereby
provides a new vacuum, around which a perturbative expansion is again
possible. Lowest-order expansion gives the Gribov-type dispersion relation
\begin{equation}
  E\left(\vec{k}^2\right)=\sqrt{\vec{k}^2+\frac{m^4}{\vec{k}^2}}.
\end{equation}
From knowledge of the Gribov mass one can deduce contributions to
the free energy $w$, the energy $e$, the (rescaled) anomaly
$A_{\mathrm{r}} = \frac{e-3p}{T^4}$ and (making use of a relation
obtained in~\cite{Kharzeev:2007wb}) the bulk viscosity $\zeta$. Results
for $m$ and $w$ are displayed in figure~\ref{fig:kldz_massfreeenergy}.

An expansion of the results in powers of $g$ is possible at least in
leading order, but this leading order term strongly deviates from the
full numerical result. This is probably not a fault of the method, but
should be regarded as yet another sign that series expansions in
thermal QCD have to be treated with great care.

Nevertheless as a matter of principle, it is a significant success that for
thermodynamic observables this procedure gives finite results
precisely at the order, $g^6$ at which ordinary perturbation theory
diverges. It also confirms the picture of thermal QCD composed of
two sectors with one being accessible perturbatively, while the other
one is genuinely nonperturbative.

\newpage

{\bf Acknowledgements}\\

\noindent K.L was supported by the Doktoratskolleg \emph{Hadronen
im Vakuum, in Kernen und Sternen} (FWF DK W1203-N08), by the
\emph{Graz Advanced School of Science} and the \emph{Paul Urban
Stipendienstiftung}.
He would like to express his thanks for the hospitality of New York
University (NYU), where most of the work contained
in~\cite{Lichtenegger:2008mh} has been done. The authors are grateful
to Reinhard Alkofer for valuable comments on this project.

\end{document}